\documentclass[runningheads]{llncs}
\newif\ifFullVersion
\FullVersiontrue
\usepackage{amsmath,amssymb}
\usepackage{marvosym}
\usepackage{url}
\usepackage[hidelinks]{hyperref}
\newcommand{\Psc}{P}
\newcommand{\Sgc}{\Sigma}
\newcommand{\Lam}{\Lambda}
\newcommand{\Pmf}{\Phi}
\newcommand{\stratA}{\mathtt{a}}
\newcommand{\stratB}{\mathtt{b}}
\newcommand{\stratC}{\mathtt{c}}
\newcommand{\appheading}[1]{\par\medskip\noindent\textit{#1.}\enspace}
\newcommand*{\fullversionurl}{}

\newcommand{\fullversionnote}{%
  \ifFullVersion\else
    \ifx\fullversionurl\empty\else
      \footnote{Full version: \url{\fullversionurl}.}%
    \fi
  \fi}
\newcommand{\fullappendix}[2]{%
  \ifFullVersion
    Appendix~\ref{app:#1}%
  \else
    \ifx\fullversionurl\empty
      Appendix~#2 of the full version%
    \else
      \href{\fullversionurl}{Appendix~#2 of the full version}%
    \fi
  \fi}

\begin{document}
\title{Blockchain Participation Games with Upfront Contribution Commitment\texorpdfstring{\thanks{\ifFullVersion Conference version accepted at IJTCS-FAW 2026 (oral presentation). \fi The author thanks several anonymous reviewers for their helpful comments. This work is partially supported by Beijing Advanced Innovation Center for Future Blockchain and Privacy Computing.}}{}}
\titlerunning{Blockchain Participation Games with Upfront Contribution Commitment}
\author{Zhenhe Zhang\inst{(\textnormal{\Letter})}}
\authorrunning{Z. Zhang}
\institute{Department of Computer Science and Technology,\\
Tsinghua University, Beijing, China\\
\email{zhenhe-z21@mails.tsinghua.edu.cn}}
\maketitle
\begin{abstract}
Blockchain participation games combine costly entry, random participation selection, and threshold success into consideration. Their retraction variants charge task-completion costs only to selected contributors. We consider committee tasks whose operators reserve costly computing or communication capacity before assignment, so preparation costs are incurred even when that capacity is not selected. We characterize all pure Nash equilibria under homogeneous selection and analyze how heterogeneous selection shapes who commits. We characterize how the homogeneous minimum-outlay reward scheme supports full contribution, where every user commits, as a Nash equilibrium. Moreover, under heterogeneous selection, requiring zero registration reward can increase the minimum outlay. Finally, when full contribution is a positive-welfare social optimum, the homogeneous zero-registration-reward design attains price of stability one; and under independent Bernoulli selection, every full-contribution Nash equilibrium is also a strong equilibrium.
\end{abstract}
\clearpage

\section{Introduction}

Blockchain consensus requires enough eligible participants to supply votes or other prescribed contributions for protocol progress. Eligibility alone supplies no contribution: an operator may enroll yet avoid the costly work on which other users rely. Blockchain participation games express this incentive problem through costly entry, random selection, and threshold success~\cite{chaidos2023blockchain}. We study its resource-commitment aspect: when productive capacity must be reserved before assignment is known, which payments make that preparation worthwhile?

The relevant time scale is a single task or committee step. Ethereum's Gasper design assigns each active validator an attestation duty every epoch~\cite{buterin2020gasper}. Protocols such as Algorand~\cite{chen2019algorand} and Ouroboros Praos~\cite{david2018ouroboros} use verifiable random functions (VRFs) for selection of leaders or committees. Private selection hides future participants until they act, limiting exposure to targeted adaptive corruption; predictable round-robin schedules reveal them in advance. We model assignment to the specified task by independent Bernoulli draws with probabilities $q_i$. A possible application is a committee service whose operators reserve computing and communication capacity before assignment. We assume this reservation binds them to execute the task if selected, and its cost is incurred even when the capacity remains unused.

Each qualified user chooses abstention ($\stratA$), bare registration ($\stratB$), or contribution commitment ($\stratC$). Registration costs $\alpha$; commitment additionally costs $\beta$, and only selected committed contributors count toward success. Success creates public value $v$ for all users. We study two separately chosen rewards: $r_1$ for registered status and an additional $r_2$ for verified selected contribution, both conditional on success. The registration reward is a candidate enrollment subsidy: it helps offset entry costs, but status-based payment also rewards users who supplied no productive capacity. Qualification fixes the eligible population before these decisions.

\paragraph{Contributions.}
\emph{Equilibrium structure.} We characterize all homogeneous pure Nash equilibria, including coexistence of abstention, bare registration, and commitment. With heterogeneous selection, we identify conditions under which more frequently selected users commit and show how registration rewards can reverse that ordering. The results distinguish public-value free-riding from reward-seeking enrollment without productive commitment (Sections~\ref{sec:hom}--\ref{sec:het}).

\emph{Reward design.} We characterize every homogeneous minimum-outlay reward scheme supporting full contribution, including when no subsidy is needed. One optimum always pays only for verified contribution and eliminates every equilibrium with bare registrants at no extra outlay. Threshold-coordination failures may remain. With heterogeneous selection, a two-user instance has a unique optimum using both rewards, cheaper than every zero-registration-reward scheme supporting full contribution (Section~\ref{sec:design}).

\emph{Welfare and stability.} Bare registration is never socially optimal: it adds real cost without improving success. We give conditions for full contribution to maximize welfare. When it is a positive-welfare optimum, the homogeneous optimal design attains price of stability one. Under arbitrary independent selection probabilities, every full-contribution Nash equilibrium is also a strong equilibrium (Section~\ref{sec:eff}).

Complete proofs and supplementary calculations appear in \fullappendix{all}{A}\ifFullVersion\else{} available on arXiv\fi.\fullversionnote

\paragraph{Related work.}
Threshold public-good and volunteer models relate shared value and pivotality to underprovision~\cite{diekmann1985volunteer,gradstein1990binary,kawagoe2023asymmetric,palfrey1984participation}. Here, random assignment separates costly preparation from the eventual use of capacity. Blockchain models study strategic investment and deviations~\cite{arnosti2022bitcoin,eyal2014majority,ferreira2022optimal,fiat2019energy,kiayias2016blockchain}, pool and delegation incentives~\cite{brunjes2020reward,gersbach2026staking,kiayias2025oceanic,kiayias2024balancing,schrijvers2016incentive}, stake-share evolution~\cite{rosu2021evolution}, consensus security~\cite{amoussou2024committee,budish2026economic}, and token adoption~\cite{cong2021tokenomics}. Repeated participation, validator entry, and monetary policy are studied in~\cite{ferraioli2025algorithmic,kiayias2025singletoken,penna2026reserve}; Airdrop Games applies participation-game methods to token allocation~\cite{georganas2025airdrop}.

Chaidos et al.~\cite{chaidos2023blockchain} provide the closest participation-game model. Their retraction variants distinguish plans to contribute if selected from plans to retract; the additional completion cost is incurred only when a contributor is selected. With zero public value ($v=0$), their heterogeneous two-action model has a selection-probability cutoff in every nontrivial pure Nash equilibrium. We retain this strategic foundation, charge commitment costs before assignment, and allow registration and verified contribution to be rewarded separately.

With homogeneous selection and $r_2=0$, their universal-payment retraction model~\cite{chaidos2023blockchain} and ours have identical ex ante payoffs when the universal reward equals $r_1$ and $\beta=q\gamma$, where $\gamma$ is their completion cost. With unequal $q_i$ and common $\gamma>0$, no common $\beta$ matches all expected costs $q_i\gamma$. Building on this shared structure, our design analysis identifies the entire optimal reward tradeoff and its equilibrium consequences, including why heterogeneous selection can make registration payments cost-effective.

\section{The Model}
\label{sec:model}

We model one round of a threshold task with random assignments. \fullappendix{model}{A.1} gives the probability tools and equilibrium-constraint derivations.

\paragraph{Users, strategies, and parameters.}
A qualification layer fixes the eligible cohort $N=\{1,\dots,n\}$ and selection probabilities $(q_i)_{i\in N}$. Each user is a distinct decision-maker controlling one qualified unit; stake, deposits, or credentials may supply Sybil resistance before this game. Users simultaneously choose $s_i\in\{\stratA,\stratB,\stratC\}$: \emph{abstention}, \emph{bare registration} without a contribution commitment, or \emph{contribution commitment}, respectively. Registration records participation; commitment reserves the required capacity. We assume a binding commitment technology executes the promised contribution if selected, with no subsequent action choice.

A profile $s$ partitions $N$ into $A(s):=\{i:s_i=\stratA\}$, $B(s):=\{i:s_i=\stratB\}$, and $C(s):=\{i:s_i=\stratC\}$. Omit $s$ when fixed and write $\lambda:=|C|$, $S_{-i}:=S\setminus\{i\}$, and $s_{-i}$ for the other users' actions. Members of $C$ are \emph{contributors}; only selected contributors count toward success. Actions $\stratA$ and $\stratB$ never count.

Parameters are registration and contribution rewards $r_1,r_2\geq0$, registration cost $\alpha>0$, preparation cost $\beta>0$, public value $v\geq0$, threshold $k\in\{1,\ldots,n\}$, and $(q_i)_{i\in N}\in(0,1)^n$. All primitives except $q_i$ are common across users; parameters and the selection law are common knowledge. Users are risk neutral and maximize ex ante expected payoff.

Before selection, each registrant in $B\cup C$ incurs $\alpha$ and each contributor additionally incurs $\beta$, both sunk regardless of assignment. These are real resource or opportunity costs, not refundable deposit principal. On success, everyone receives $v$, every registrant receives $r_1$, and each selected contributor additionally receives $r_2$; failure pays none of these benefits. There is no additional execution cost after selection.

\paragraph{Selection and success.}
After actions and costs are fixed, mutually independent potential assignments $X_i\sim\mathrm{Bernoulli}(q_i)$ are drawn from an exogenous, action-independent law. For $S\subseteq N$, write $X_S:=\sum_{i\in S}X_i$. The task succeeds exactly when $X_C\geq k$, after which rewards are paid. For integers $j,x$, define
\begin{align}
\Sgc(j,S)&:=\Pr[X_S=j]
=\sum_{\substack{T\subseteq S\\ |T|=j}}
  \prod_{i\in T}q_i\prod_{h\in S\setminus T}(1-q_h), \label{eq:sigma-def}\\
\Psc(x,S)&:=\Pr[X_S\geq x]=\sum_{j=x}^{|S|}\Sgc(j,S). \label{eq:p-def}
\end{align}
We set $\Sgc(j,S)=0$ for $j<0$ or $j>|S|$, $\Psc(x,S)=1$ for $x\leq0$, and $\Psc(x,S)=0$ for $x>|S|$; impossible-index sums are empty.

Under homogeneous selection $q_i=q\in(0,1)$, probabilities depend only on $\ell=|S|$:
\begin{equation*}
\Sgc(j,\ell):=\binom{\ell}{j}q^j(1-q)^{\ell-j},
\qquad
\Psc(x,\ell):=\sum_{j=x}^{\ell}\Sgc(j,\ell),
\end{equation*}
with the same boundary conventions. Conditioning on $X_i$ gives, for every $i\notin S$, the recursion
\begin{equation}\label{eq:rec}
\Psc(x,S\cup\{i\}) = q_i\,\Psc(x-1,S) + (1-q_i)\,\Psc(x,S),
\end{equation}
and $\Sgc(j,S\cup\{i\})=q_i\Sgc(j-1,S)+(1-q_i)\Sgc(j,S)$. Hence $\Psc(x,S\cup\{i\})-\Psc(x,S)=q_i\Sgc(x-1,S)$ and $\Psc(x-1,S)-\Psc(x,S)=\Sgc(x-1,S)$.

\paragraph{Payoffs.}
Fix $s_{-i}$ and let $C_{-i}:=\{j\neq i:s_j=\stratC\}$. Expected payoffs are
\begin{align}
u_i(\stratA, s_{-i}) &= v\,\Psc(k,C_{-i}), \label{eq:ua}\\
u_i(\stratB, s_{-i}) &= (r_1+v)\,\Psc(k,C_{-i})-\alpha, \notag\\
u_i(\stratC, s_{-i}) &= -\alpha-\beta
 +q_i(r_1+r_2+v)\,\Psc(k-1,C_{-i}) \notag\\
&\quad +(1-q_i)(r_1+v)\,\Psc(k,C_{-i}). \label{eq:uc}
\end{align}
In \eqref{eq:uc}, a selected user needs $k-1$ other selected contributors; an unselected user needs $k$.
A \emph{pure Nash equilibrium} satisfies $u_i(s_i,s_{-i})\geq u_i(y,s_{-i})$ for every user $i$ and action $y$; it is \emph{strict} if all inequalities for $y\neq s_i$ are strict. Positive success probability is equivalent to $|C|\geq k$, since every $q_i\in(0,1)$.

\paragraph{Two marginal-value functions.}
For a pool $S$ of other contributors, define
\begin{equation}\label{eq:lam-phi}
\begin{split}
\Lam(S) &:= r_2\,\Psc(k-1,S) + v\,\Sgc(k-1,S),\\
\Pmf(S) &:= r_2\,\Psc(k-1,S) + (r_1+v)\,\Sgc(k-1,S) \;=\; \Lam(S) + r_1\Sgc(k-1,S).
\end{split}
\end{equation}
Commitment gains over abstention and bare registration are respectively $-\alpha-\beta+q_i\Lam(S)+r_1\Psc(k,S\cup\{i\})$ and $-\beta+q_i\Pmf(S)$. A selected contributor is pivotal when exactly $k-1$ other contributors are selected; the additional term $r_1\Sgc(k-1,S)$ captures the resulting registration reward. Unlike the tail, this pivotal mass need not increase with the pool.

\begin{lemma}[Equilibrium constraints]\label{lem:constraints}
A profile $(A,B,C)$ is a pure Nash equilibrium if and only if every displayed inequality holds for every user in its indicated set:
\begin{align}
\alpha &\geq r_1 \Psc(k,C) && i \in A \label{eq:I}\\
\alpha &\leq r_1 \Psc(k,C) && i \in B \label{eq:II}\\
\alpha + \beta &\geq q_i\,\Lam(C) + r_1 \Psc(k, C\cup\{i\}) && i \in A \label{eq:III}\\
\beta &\geq q_i\,\Pmf(C) && i \in B \label{eq:IV}\\
\alpha + \beta &\leq q_i\,\Lam(C_{-i}) + r_1 \Psc(k,C) && i \in C \label{eq:V}\\
\beta &\leq q_i\,\Pmf(C_{-i}) && i \in C \label{eq:VI}
\end{align}
Empty-set conditions are vacuous. The rows exclude deviations $\stratA\to\stratB$, $\stratB\to\stratA$, $\stratA\to\stratC$, $\stratB\to\stratC$, $\stratC\to\stratA$, and $\stratC\to\stratB$, respectively. In equilibrium, $C\neq\emptyset$ implies $\lambda\geq k$; otherwise contributors strictly prefer abstention to $-\alpha-\beta$.
\end{lemma}
Conditions \eqref{eq:I}--\eqref{eq:II} allow $A,B\neq\emptyset$ only when $\alpha=r_1\Psc(k,C)$, the equality driving three-action coexistence.

\section{Homogeneous Selection Probabilities}
\label{sec:hom}

We characterize all pure equilibria for $q_i=q\in(0,1)$. For $|S|=\ell$, write $\Lam(\ell):=\Lam(S)$ and $\Pmf(\ell):=\Pmf(S)$, extending the pool-size notation for $\Psc$ and $\Sgc$; $\lambda=|C|$. Complete proofs, reward intervals, and example calculations appear in \fullappendix{hom}{A.2}.

\subsection{Single-strategy profiles}

\begin{proposition}[Single-strategy equilibria]\label{prop:single}
All-abstention ($B=C=\emptyset$) is an equilibrium for $k\geq2$; for $k=1$ it is an equilibrium iff $\alpha+\beta\geq q(r_1+r_2+v)$. The profile in which all users choose bare registration ($C=\emptyset$, $B=N$) is never an equilibrium. The full-contribution profile ($C = N$) is a Nash equilibrium iff
\begin{equation}\label{eq:allC}
\alpha + \beta \;\leq\; q\Lam(n-1) + r_1 \Psc(k,n)
\qquad\text{and}\qquad
\beta \;\leq\; q\Pmf(n-1).
\end{equation}
As $n \to \infty$ with all other parameters fixed, the right-hand sides of these inequalities converge to $r_1+qr_2$ and $qr_2$, respectively.
\end{proposition}

As the pool grows, success becomes nearly certain and individual pivotality vanishes. For $k\geq2$, however, one commitment cannot reach the threshold: success-contingent rewards sustain contribution after coordinated entry but cannot eliminate all-abstention. Coordinating entry or compensating preparation before success could address this failure when participation creates positive welfare.

No equilibrium has $C=\emptyset$ and $B\neq\emptyset$, since bare registration would require $\alpha\leq r_1\Psc(k,\emptyset)=0$. Thus the following $\stratA\stratC$, $\stratB\stratC$, and $\stratA\stratB\stratC$ cases exhaust multi-action equilibria.

\subsection{Abstention--commitment equilibria}
\label{sec:ac}

An $\stratA\stratC$ equilibrium has $B=\emptyset$ and $k\leq\lambda\leq n-1$; its window for $\alpha+\beta$ is
\begin{equation}\label{eq:ac-window}
q\Lam(\lambda) + r_1\Psc(k,\lambda+1) \;\leq\; \alpha+\beta \;\leq\; q\Lam(\lambda-1) + r_1\Psc(k,\lambda).
\end{equation}

\begin{theorem}[$\stratA\stratC$ equilibria]\label{thm:ac}
Fix $k \leq \lambda \leq n-1$. A profile with $B=\emptyset$ and $|C|=\lambda$ is a Nash equilibrium if and only if the window \eqref{eq:ac-window} contains $\alpha+\beta$ and
\begin{equation}\label{eq:ac-side}
\alpha \geq r_1\Psc(k,\lambda),
\qquad
\beta \leq q\Pmf(\lambda-1).
\end{equation}
For fixed $(r_1,r_2,v,q,k,\lambda)$, there exist positive costs $(\alpha,\beta)$ supporting such a profile if and only if $\Lam(\lambda-1)>0$ and
\begin{equation}\label{eq:wac}
q\,(r_1+r_2+v)\,\Sgc(k-2,\lambda-1) \;\leq\; \big(qv - (1-q)r_1\big)\,\Sgc(k-1,\lambda-1)
\end{equation}
(with $\Sgc(-1,\cdot)=0$). Equation~\eqref{eq:wac} tests window nonemptiness, not membership of the given $\alpha+\beta$. In particular:
\begin{itemize}
\item[(a)] If $qv < (1-q)r_1$ (in particular whenever $v = 0 < r_1$), no $\stratA\stratC$ equilibrium exists for any $\lambda$; the same holds at the boundary $qv = (1-q)r_1$ when $k \geq 2$: \eqref{eq:wac} fails unless all rewards and $v$ vanish, in which case $\Lam(\lambda-1)=0$ precludes positive costs.
\item[(b)] If $qv > (1-q)r_1$ and $k \geq 2$, positive supporting costs exist iff
\begin{equation}\label{eq:ac-lambda}
\lambda \;\geq\; \lambda_{\stratA\stratC} \;:=\; k - 1 + (k-1)\,\frac{(1-q)(r_1+r_2+v)}{qv - (1-q)r_1}.
\end{equation}
\item[(c)] If $k = 1$, positive supporting costs exist iff $qv \geq (1-q)r_1$ and $\Lam(\lambda-1)>0$; in particular they exist for every $\lambda \geq 1$ when the first inequality is strict.
\end{itemize}
\end{theorem}

\begin{remark}[Meaning of the $\stratA\stratC$ threshold]\label{rem:correction}
For $k\geq2$, coexistence requires $qv-(1-q)r_1>0$: public value sustains commitment, while registration rewards attract abstainers. Raising $r_1$ or $r_2$ raises $\lambda_{\stratA\stratC}$; raising $v$ weakly lowers it. A larger $r_2$ can induce abstainers to commit, ending coexistence. This threshold describes partial participation, not full-contribution implementation. For $k=1$, nonnegative pivotal advantage and $\Lam(\lambda-1)>0$ suffice, without a size threshold.
\end{remark}

\subsection{Bare-registration--commitment equilibria}
\label{sec:bc-hom}

The lower and upper bounds of the $\stratB\stratC$ preparation-cost window rule out profitable deviations from bare registration to commitment and from commitment to bare registration, respectively.

\begin{theorem}[$\stratB\stratC$ equilibria]\label{thm:bc-hom}
A profile with $A=\emptyset$, $B,C\neq\emptyset$, and $|C|=\lambda\geq k$ is a Nash equilibrium if and only if
\begin{equation}\label{eq:bc-exact}
\begin{split}
q\Pmf(\lambda)&\leq\beta\leq q\Pmf(\lambda-1),\qquad
\alpha\leq r_1\Psc(k,\lambda),\\
\alpha+\beta&\leq q\Lam(\lambda-1)+r_1\Psc(k,\lambda).
\end{split}
\end{equation}
The preparation-cost interval in \eqref{eq:bc-exact} is nonempty exactly when $\Pmf(\lambda)\leq\Pmf(\lambda-1)$. For $k=1$ this always holds. For $k\geq2$ and $r_1+v>0$, it is equivalent to
\begin{equation}\label{eq:bc-lambda}
\lambda \;\geq\; \lambda_{\stratB\stratC} \;:=\; k - 1 + (k-1)\,\frac{(1-q)(r_1+r_2+v)}{q\,(r_1+v)},
\end{equation}
whereas for $k\geq2$ and $r_1+v=0$ it holds exactly when $r_2=0$. 
\end{theorem}

The thresholds describe different incentives. In $\stratA\stratC$, public value offsets the registration reward. In $\stratB\stratC$, a larger pool can dilute commitment's incremental value, allowing users to retain registration rewards without paying $\beta$. Equation~\eqref{eq:bc-exact} additionally tests the given costs.

\subsection{Three-action coexistence equilibria}
\label{sec:abc}

\begin{theorem}[$\stratA\stratB\stratC$ equilibria]\label{thm:abc-hom}
A profile with $A, B, C$ all nonempty and $|C| = \lambda \geq k$ is a Nash equilibrium if and only if
\begin{equation}\label{eq:abc-cond}
\alpha = r_1\Psc(k,\lambda)
\qquad\text{and}\qquad
q\,\Pmf(\lambda) \;\leq\; \beta \;\leq\; q\,\Lam(\lambda-1).
\end{equation}
For fixed $(r_1,r_2,v,q,k,\lambda)$, positive supporting costs exist iff $r_1>0$ and \eqref{eq:wac} holds. For $r_1>0$, this requires $qv>(1-q)r_1$ and $\lambda\geq\lambda_{\stratA\stratC}$ when $k\geq2$, or $qv\geq(1-q)r_1$ when $k=1$. Every such equilibrium has $r_1>0$ and is non-strict: all users in $A\cup B$ are indifferent between $\stratA$ and $\stratB$.

For a fixed three-action profile and all other parameters including $\beta$, a positive supporting $\alpha$ exists iff $r_1>0$ and $q\Pmf(\lambda)\leq\beta\leq q\Lam(\lambda-1)$; it is uniquely $r_1\Psc(k,\lambda)$. Increasing it violates \eqref{eq:II}; decreasing it violates \eqref{eq:I}.
\end{theorem}

\begin{example}\label{ex:abc}
Take $k=2$, $q=1/2$, $r_1=r_2=1$, $v=3$, $n=6$, $\lambda=4$, and $|A|=|B|=1$. Then $\alpha=\Psc(2,4)=11/16$ and every $\beta\in[31/32,1]$ yield three-way coexistence. Holding others' actions fixed, raising $\alpha$ makes a bare registrant prefer abstention; lowering it makes an abstainer prefer bare registration. The profile fixes one registration cost but allows a range of preparation costs.
\end{example}

\section{Heterogeneous Selection Probabilities}
\label{sec:het}

Now allow arbitrary $(q_i)_{i\in N}\in(0,1)^n$ under the independent draws of Section~\ref{sec:model}. Complete proofs, reward intervals, and example calculations appear in \fullappendix{het}{A.3}.

\begin{definition}[Registration dominance, RD]\label{def:rd}
User $i$ satisfies \emph{registration dominance} if $(1-q_i)\,r_1 \geq q_i\,v$, and satisfies \emph{strict registration dominance} if the inequality is strict. We say the population satisfies RD or strict RD when every user satisfies the corresponding condition.
\end{definition}

RD compares the registration reward weighted by nonselection with public value weighted by selection, the pivotal balance in Theorem~\ref{thm:ac}. It orders abstainers below contributors; boundary incentives then determine whether a candidate cutoff is an equilibrium. The cutoff results assume RD for $k\geq2$ and strict RD for $k=1$.

\subsection{Abstention--commitment equilibria}

The next lemmas exclude equal-probability splits, order abstainers below contributors, and reduce incentive checks to boundary users.

\begin{lemma}[No equal-probability abstainer--contributor split]\label{lem:equal}
Assume \hyperref[def:rd]{RD} when $k\geq2$ and strict \hyperref[def:rd]{RD} when $k=1$. In any equilibrium, two users $e, f$ with $q_e = q_f$ cannot satisfy $e \in A$ and $f \in C$.
\end{lemma}

\begin{lemma}[No more-selectable abstainer]\label{lem:dom}
Assume \hyperref[def:rd]{RD} when $k\geq2$ and strict \hyperref[def:rd]{RD} when $k=1$. If $e\in A$ and $f\in C$ in equilibrium, then $q_e<q_f$. Consequently, in an $\stratA\stratC$ equilibrium, every user more selectable than a contributor also commits to contribute.
\end{lemma}

\begin{lemma}[Monotonicity of staying incentives]\label{lem:monoC}
Let $e, f \in C$ with $q_f > q_e$. If $e$ satisfies \eqref{eq:V} and \eqref{eq:VI}, so does $f$. Likewise, let $e, f \in A$ with $q_f > q_e$: if $f$ satisfies
\eqref{eq:III}, so does $e$.
\end{lemma}

\begin{theorem}[Contribution-threshold theorem]\label{thm:threshold}
Assume \hyperref[def:rd]{RD} when $k\geq2$ and strict \hyperref[def:rd]{RD} when $k=1$. Every $\stratA\stratC$ equilibrium has $C=\{i:q_i\geq q_d\}$ for some threshold $q_d$, with $|C|\geq k$; all remaining users abstain. A candidate $\stratA\stratC$ cutoff with $k\leq|C|\leq n-1$ is an equilibrium iff \eqref{eq:I} holds, its least-selectable contributor satisfies \eqref{eq:V}--\eqref{eq:VI}, and its most-selectable abstainer satisfies \eqref{eq:III}.
\end{theorem}

Sorting by $q_i$ leaves at most $n-k$ nontrivial cutoffs without splitting ties. The lowest contributor tests withdrawal and the highest abstainer tests entry; Lemma~\ref{lem:constraints} separately covers all-abstention and full contribution. For $k=1$, weak RD at equality can allow equal-$q_i$ users in both $A$ and $C$; \fullappendix{het}{A.3} gives this boundary case and a strict-RD example.

\subsection{Bare-registration--commitment equilibria without a cutoff}
\label{sec:bc-het}

With bare registration, contributors need not have the highest selection probabilities; the next result characterizes any candidate $C$ directly through boundary conditions.

\begin{theorem}[$\stratB\stratC$ boundary conditions]\label{thm:bc-het}
Let $A=\emptyset$, let $B,C$ be nonempty, and let $|C|\geq k$. Choose any least-selectable contributor $\underline c\in\arg\min_{i\in C}q_i$ and any most-selectable bare registrant $\bar b\in\arg\max_{j\in B}q_j$. The profile is a Nash equilibrium if and only if:
\begin{align}
\beta &\;\geq\; q_{\bar b}\,\Pmf(C), \label{eq:W1}\\
\beta &\;\leq\; q_{\underline c}\,\Pmf(C_{-\underline c}), \label{eq:W2}\\
\alpha &\;\leq\; r_1\Psc(k,C), \label{eq:W3}\\
\alpha + \beta &\;\leq\; q_{\underline c}\,\Lam(C_{-\underline c}) + r_1\Psc(k,C). \label{eq:W4}
\end{align}
These conditions reduce to Theorem~\ref{thm:bc-hom} when $q_i\equiv q$.
\end{theorem}

The least-selectable contributor and most-selectable bare registrant have the tightest withdrawal and entry constraints, respectively. Bounds on $r_2$ subject to \eqref{eq:W3} are derived in \fullappendix{het}{A.3}.

\begin{example}[Ordering violation in a $\stratB\stratC$ equilibrium]\label{ex:wrong-bc}
Let $n=2$, $k=1$, $v=0$, $r_1=1$, $r_2=0.1$, $(q_1,q_2)=(0.4,0.5)$, $\alpha=0.03$, $\beta=0.4$, $C=\{1\}$, and $B=\{2\}$. Then $\Pmf(\{1\})=0.7$ and $\Pmf(\emptyset)=1.1$ satisfy \eqref{eq:W1}--\eqref{eq:W4}: the more-selectable user free-rides through bare registration on the less-selectable contributor. Proposition~\ref{prop:restore} gives a sufficient condition excluding this reversal.
\end{example}

A bare registrant receives $r_1$ on success. Commitment adds the contribution reward and, conditional on selection, the pivotal gain $(r_1+v)\Sgc(k-1,C)$. A larger pool can reduce this gain as success without the user becomes likelier:
\begin{equation}\label{eq:phi-diff}
\begin{split}
\Pmf(C)-\Pmf(C_{-i})
={}&q_i\big[(r_1+r_2+v)\Sgc(k-2,C_{-i})\\
&\qquad -(r_1+v)\Sgc(k-1,C_{-i})\big].
\end{split}
\end{equation}
A negative difference permits reversed ordering. Under \emph{pivotal non-dilution}, adding back a contributor does not reduce the gain over bare registration:

\begin{proposition}[Pivotal non-dilution restores ordering]\label{prop:restore}
Suppose the contributor set satisfies $\Pmf(C) \geq \Pmf(C_{-i})$ for every $i \in C$, equivalently $(r_1+r_2+v)\Sgc(k-2,C_{-i}) \geq (r_1+v)\Sgc(k-1,C_{-i})$. Then any $\stratB\stratC$ equilibrium with contributor set $C$ has $\min_{i \in C} q_i \geq \max_{j \in B} q_j$; the inequality is strict whenever
\eqref{eq:phi-diff} is strictly positive for the least-selectable contributor
$\underline c$ (chosen from $\arg\min_{i\in C}q_i$), or all selection probabilities are distinct.
\end{proposition}

For $k = 1$, \eqref{eq:phi-diff} equals $-q_i(r_1+v)\Sgc(0,C_{-i})<0$ in any $\stratB\stratC$ equilibrium, so pivotal non-dilution fails; Example~\ref{ex:wrong-bc} shows that an ordering violation can occur.

\subsection{Three-action equilibria under heterogeneous selection}

\begin{theorem}[Heterogeneous $\stratA\stratB\stratC$ equilibria]\label{thm:abc-het}
Fix a profile with $A,B,C$ all nonempty, contributor set $C$, and $|C|\geq k$. Choose any $\underline c\in\arg\min_{i\in C}q_i$ and any $\bar d\in\arg\max_{j\in A\cup B}q_j$. The profile is a Nash equilibrium if and only if
\begin{equation}\label{eq:abc-het}
\alpha \;=\; r_1\Psc(k,C)
\qquad\text{and}\qquad
q_{\bar d}\,\Pmf(C) \;\leq\; \beta \;\leq\; q_{\underline c}\,\Lam(C_{-\underline c}).
\end{equation}
These equilibria are never strict. A fixed profile and fixed remaining parameters admit at most one supporting registration cost, namely $\alpha=r_1\Psc(k,C)$. Existence is strictly stronger than consistency of the $\stratB\stratC$ window \eqref{eq:W1}--\eqref{eq:W2}.
\end{theorem}

The window narrows at both ends: its lower endpoint uses the largest $q_i$ in $A\cup B$, and its upper endpoint uses $\Lam\leq\Pmf$.

\begin{example}[Ordering violation in a three-action equilibrium]\label{ex:wrong-abc}
Let $n=3$, $k=1$, $r_1=r_2=1$, $v=10$, with $q_1=0.1$ (abstains), $q_2=0.55$ (chooses bare registration), and $q_3=0.5$ (commits). Set $\alpha=r_1\Psc(1,\{3\})=0.5$ and $\beta=4$. Then $q_2\Pmf(\{3\})=3.575\leq4\leq5.5=q_3\Lam(\emptyset)$ satisfies \eqref{eq:abc-het}: the most-selectable user chooses bare registration while a less-selectable user commits.
\end{example}

\section{Optimal Reward Design}
\label{sec:design}

We design success-contingent rewards $(r_1,r_2)$ that support full contribution $(B=\emptyset,C=N)$ as a Nash equilibrium at minimum expected protocol outlay. At a profile $(B,C)$, that outlay is
\begin{equation}\label{eq:outlay}
\operatorname{Out}(r_1,r_2;B,C) \;=\; r_1\,(|B|+|C|)\,\Psc(k,C) \;+\; r_2 \sum_{i \in C} q_i\,\Psc(k-1,C_{-i}).
\end{equation}
The two terms pay for registered status and selected contribution, respectively. Public value $v$ is not protocol outlay: pivotal public value reduces the subsidy needed to sustain contribution.

\fullappendix{design}{A.4} contains the complete proofs of Theorem~\ref{thm:design}, Corollary~\ref{cor:nofree}, and the heterogeneous-design claims in Remark~\ref{rem:hetero-design}.

For real $x$, write $x^{+}:=\max\{x,0\}$.

\begin{theorem}[Minimum-outlay reward schemes]\label{thm:design}
Under homogeneous selection probabilities, let $\Gamma_1 := \alpha+\beta - qv\,\Sgc(k-1,n-1)$ and $\Gamma_2 := \beta - qv\,\Sgc(k-1,n-1)$. These are the total and incremental private costs, respectively, net of pivotal public value. If $\Gamma_1 \leq 0$, the unique minimum-outlay scheme is $(r_1,r_2)=(0,0)$, with $\operatorname{Out}^{*}=0$, and all-$\stratC$ is a Nash equilibrium. If $\Gamma_1 > 0$, the minimum expected protocol outlay subject to all-$\stratC$ being a Nash equilibrium is $\operatorname{Out}^{*} = n\,\Gamma_1$, attained exactly by the segment of reward schemes
\begin{equation}\label{eq:segment}
\Psc(k,n)\, r_1 + q\,\Psc(k-1,n-1)\, r_2 = \Gamma_1,
\qquad
0 \leq r_1 \leq \min\Big\{\frac{\alpha}{\Psc(k,n-1)},\;
\frac{\Gamma_1}{\Psc(k,n)}\Big\}
\end{equation}
(with the first term of the minimum read as $+\infty$ when $\Psc(k,n-1) = 0$). In both cases the zero-registration-reward scheme
\begin{equation}\label{eq:opt-endpoint}
r_1^{*} = 0,
\qquad
r_2^{*} = \frac{\Gamma_1^{+}}{q\,\Psc(k-1,n-1)}
\end{equation}
is optimal; when $\Gamma_1>0$, it is the only point of the segment with zero registration reward.
\end{theorem}

\begin{proof}
The constraint against abstention in \eqref{eq:allC} gives $\operatorname{Out}/n\geq\Gamma_1$. When $\Gamma_1>0$, imposing equality makes the constraint against bare registration equivalent to $r_1\Psc(k,n-1)\leq\alpha$. Together with nonnegative rewards, these are exactly \eqref{eq:segment}; its $r_1=0$ endpoint is feasible. When $\Gamma_1\leq0$, zero rewards satisfy both constraints and uniquely minimize outlay, whose coefficients are positive.\qed
\end{proof}
Moving along the segment reallocates the same outlay between registration and selected contribution. Its $r_1=0$ endpoint also changes the equilibrium set:

\begin{corollary}[Eliminating bare registration]\label{cor:nofree}
Under homogeneous selection, at the optimal endpoint $(r_1^{*}, r_2^{*})$ the strategy $\stratB$ is strictly dominated by $\stratA$, so every equilibrium has $B = \emptyset$: no $\stratB\stratC$ or $\stratA\stratB\stratC$ equilibrium survives. If $v = 0$ and $k \geq 2$, no $\stratA\stratC$ equilibrium exists either, so the only equilibria are all-$\stratA$ and all-$\stratC$, and full contribution is the \emph{unique equilibrium with positive success probability}. For $k = 1$, holding $r_1=0$, \emph{any} $r_2 > r_2^{*}$ makes all-$\stratC$ the \emph{unique} equilibrium.
\end{corollary}

For $k\geq2$, all-$\stratA$ remains because one deviator cannot activate a success-contingent payoff. When $v>0$, partial $\stratA\stratC$ equilibria may also survive through public-value free-riding. For $k=1$, the strict contribution-reward choice in Corollary~\ref{cor:nofree} eliminates both forms of abstention.

\begin{remark}[Reward design under heterogeneous selection]\label{rem:hetero-design}
For arbitrary $(q_i)$ the problem is a two-variable linear program. At full contribution, $a_i:=q_i\Psc(k-1,N_{-i})$ is the probability that user $i$ is selected and the round succeeds. With $\tilde Q:=\sum_i a_i$, balanced exposure $a_i=\tilde Q/n$ for every $i$ guarantees an optimum with $r_1=0$: shifting payment to $r_2$ at equal outlay preserves the abstention constraints and weakly relaxes the constraints against bare registration. When $k<n$, balanced exposure is equivalent to homogeneous selection; when $k=n$, it holds for arbitrary selection probabilities because $a_i=\prod_{j\in N}q_j$. When exposure is unbalanced, a common registration payment may incentivize a low-$a_i$ user at lower total outlay than the contribution reward. For $n=2$, $k=1$, $(q_1,q_2)=(1/3,2/3)$, $\alpha=2$, $\beta=3$, and $v=0$, the unique optimum is $(r_1,r_2)=(3,8)$: its outlay is $38/3$, whereas the best scheme with $r_1=0$ costs $15$.
\end{remark}

\section{Social Efficiency}
\label{sec:eff}

Welfare and price-of-stability results assume homogeneous selection; strong equilibrium allows arbitrary independent probabilities. \fullappendix{eff}{A.5} contains all proofs. Including users and the protocol payment account makes rewards cancel as transfers:
\begin{equation}\label{eq:welfare}
W(s) \;=\; nv\,\Psc(k,C) \;-\; \alpha\,(|B|+|C|) \;-\; \beta\,|C|.
\end{equation}
Let $\mathcal E$ be the pure Nash equilibria and $\mathrm{OPT}:=\max_sW(s)$. For $\mathcal E\neq\emptyset$ and $\mathrm{OPT}>0$, define $\mathrm{PoS}:=\mathrm{OPT}/\max_{s\in\mathcal E}W(s)$ and $\mathrm{PoA}:=\mathrm{OPT}/\min_{s\in\mathcal E}W(s)$; either ratio is $+\infty$ when its denominator is nonpositive.

\begin{proposition}[Structure of the welfare optimum]\label{prop:opt}
\begin{enumerate}
\item[(i)] Every profile with $B \neq \emptyset$ has strictly lower welfare than the profile obtained by moving $B$ to $A$, so every welfare-maximizing profile has $B = \emptyset$.
\item[(ii)] Writing $W(\lambda) := nv\,\Psc(k,\lambda) - \lambda(\alpha+\beta)$ for the welfare of a profile with $\lambda$ contributors and no bare registrants, $W(\lambda) - W(\lambda-1) = nvq\,\Sgc(k-1,\lambda-1) - (\alpha+\beta)$; hence full contribution is socially optimal whenever $nvq\,\Sgc(k-1,\lambda-1) \geq \alpha+\beta$ for all $k \leq \lambda \leq n$ and $nv\,\Psc(k,n) \geq n(\alpha+\beta)$.
\end{enumerate}
\end{proposition}

Bare registration can be privately attractive when $r_1>0$, but removing it saves $\alpha$ without affecting success; setting $r_1=0$ eliminates it in equilibrium. The marginal conditions make expansions from $k$ through $n$ weakly welfare-improving, while the level condition compares full contribution with all subthreshold profiles, including the zero-welfare all-abstention profile.

\begin{theorem}[Price of stability under optimal design]\label{thm:pos}
The scheme \eqref{eq:opt-endpoint} supports full contribution as a Nash equilibrium. If that profile is a positive-welfare social optimum, the game's price of stability is $1$.
\end{theorem}
For $k\geq2$, all-abstention remains a zero-welfare equilibrium: under this theorem's premise, $\mathrm{PoS}=1$ but $\mathrm{PoA}=+\infty$. For $k=1$, Corollary~\ref{cor:nofree}'s strict reward increase makes full contribution unique. A \emph{strong equilibrium} permits no coalition whose members all gain strictly from a joint deviation.

\begin{theorem}[Strong full contribution]\label{thm:strong}
For arbitrary independent selection probabilities $(q_i)\in(0,1)^n$, every all-$\stratC$ Nash equilibrium is a strong equilibrium.
\end{theorem}

\begin{proof}
For any fixed action, a user's payoff is nondecreasing in the set of other contributors. A coalition deviating from full contribution can only shrink that set, so each member's deviating payoff is at most the payoff from the same action against all other users contributing. The Nash condition bounds this by the member's original payoff.\qed
\end{proof}

\section{Conclusion}

This paper studies how participation changes when costly readiness must be chosen before random selection. The model distinguishes registration from productive readiness and identifies two forms of free-riding: users may abstain and rely on public value created by others, or register for rewards without reserving productive capacity. We completely characterize pure-strategy equilibria under homogeneous selection, including coexistence of all three behaviors, and show how heterogeneous selection reshapes who commits. In particular, a greater likelihood of selection need not produce a greater willingness to commit when registration is rewarded separately.

Under homogeneous selection, the reward-design analysis characterizes the entire family of minimum-outlay schemes that support full contribution. This family always includes a design with no registration reward, so any required subsidy can be directed toward verified contribution without increasing expected outlay. That design removes every equilibrium containing bare registrants. All-abstention may persist through threshold coordination, and partial contribution through public-value free-riding. Under heterogeneous selection, the two rewards need not remain interchangeable: in our two-user instance, the unique minimum-outlay design uses both rewards and is strictly cheaper than every design with no registration reward supporting full contribution.

The welfare analysis shows why registration and productive contribution must be distinguished. Bare registration may be privately attractive, but it never increases the success probability and incurs a real cost, so it cannot be socially optimal. When full contribution is a positive-welfare social optimum, the homogeneous minimum-outlay design with no registration reward supports that outcome at equilibrium and attains price of stability one. Under independent selection with arbitrary user probabilities, any full-contribution Nash equilibrium admits no profitable coalitional deviation and hence is a strong equilibrium. Together, the equilibrium conditions identify the deviations that rewards must deter, the design analysis determines the least expenditure needed to support full contribution, and the welfare comparison determines when this target is socially desirable.

\paragraph{Future work.} Task-level measurements of assignment rates and preparation costs would enable protocol calibration. Mixed strategies would extend the equilibrium analysis to randomized participation, while adjustment dynamics could explain which outcomes emerge when equilibria coexist. Correlated selection would require a joint model of assignments, while endogenous entry and the creation or joint control of multiple identities would require participation rewards to be studied together with qualification and Sybil resistance. Endogenous token prices would link the real value of rewards to participation, and post-selection execution costs would distinguish the cost of readiness from the cost of performing the task. Lockups, compounding, reserve depletion, repeated participation, and history-dependent slashing would add intertemporal state and connect current reward design to longer-run protocol incentives. The repeated-participation and reserve models in \cite{ferraioli2025algorithmic,penna2026reserve} provide starting points for these intertemporal extensions.

\begin{credits}
\subsubsection{Disclosure of Interests.}
The author has no competing interests to declare that are relevant to the content of this article.
\end{credits}
\clearpage
\bibliographystyle{splncs04}
\begingroup
\let\url\nolinkurl
\bibliography{refs}
\endgroup

\clearpage
\appendix
\renewcommand{\theHsection}{appendix.\Alph{section}}
\section{Complete Proofs Deferred from the Main Text}
\label{app:proofs}\label{app:all}

This appendix gives the deferred proofs and example calculations in the order of the main text, including the strict-\hyperref[def:rd]{RD} instance and the weak-\hyperref[def:rd]{RD} boundary counterexample.

\subsection{Probability identities and equilibrium constraints}
\label{app:model}

For homogeneous selection, the binomial masses satisfy, for $k\geq2$ and $\ell\geq k-1$,
\begin{equation}\label{eq:binom-ratio}
\frac{\Sgc(k-1,\ell)}{\Sgc(k-2,\ell)}
=\frac{\ell-k+2}{k-1}\cdot\frac{q}{1-q}.
\end{equation}

\begin{proof}[Lemma~\ref{lem:constraints}]
Fix $i$. If $i\notin C$, its other-contributor set is $C$. Direct subtraction in
\eqref{eq:ua}--\eqref{eq:uc}, followed by \eqref{eq:rec}, gives
\begin{align*}
u_i(\stratB,s_{-i})-u_i(\stratA,s_{-i})
  &=r_1\Psc(k,C)-\alpha,\\
u_i(\stratC,s_{-i})-u_i(\stratA,s_{-i})
  &=-\alpha-\beta+q_i\Lam(C)+r_1\Psc(k,C\cup\{i\}),\\
u_i(\stratC,s_{-i})-u_i(\stratB,s_{-i})
  &=-\beta+q_i\Pmf(C).
\end{align*}
For an abstainer, nonpositivity of the first two differences is precisely conditions \eqref{eq:I} and \eqref{eq:III}; for a bare registrant, nonnegativity of the first difference and nonpositivity of the third difference are precisely conditions \eqref{eq:II} and
\eqref{eq:IV}. If $i\in C$, put
$D:=C_{-i}$. The same subtraction gives
\begin{align*}
u_i(\stratC,s_{-i})-u_i(\stratA,s_{-i})
  &=-\alpha-\beta+q_i\Lam(D)+r_1\Psc(k,C),\\
u_i(\stratC,s_{-i})-u_i(\stratB,s_{-i})
  &=-\beta+q_i\Pmf(D),
\end{align*}
which are nonnegative exactly under \eqref{eq:V}--\eqref{eq:VI}. These comparisons exhaust the two alternative actions of every user, proving necessity and sufficiency. Finally, if $0<|C|<k$, success remains impossible when any contributor deviates to $\stratA$, so that deviation raises the user's payoff from $-\alpha-\beta$ to zero. Hence a nonempty equilibrium contributor set has $|C|\geq k$.
\qed
\end{proof}

\subsection{Homogeneous selection}
\label{app:hom}

\begin{proof}[Proposition~\ref{prop:single}]
For all-abstention, $B=C=\emptyset$, \eqref{eq:I} is automatic and \eqref{eq:III} reads $\alpha+\beta\geq q\Lam(\emptyset)+r_1\Psc(k,\{i\})$. For $k\geq2$ both terms on the right vanish. For $k=1$, $\Lam(\emptyset)=r_2+v$ and $\Psc(1,\{i\})=q$, giving $\alpha+\beta\geq q(r_1+r_2+v)$.

If all users choose bare registration, $B=N$ and $C=\emptyset$, \eqref{eq:II} requires $\alpha\leq r_1\Psc(k,\emptyset)=0$, contrary to $\alpha>0$.

For full contribution, $C=N$, only \eqref{eq:V}--\eqref{eq:VI} apply and homogeneity makes them exactly \eqref{eq:allC}, proving the equivalence. For fixed $k$ and $q\in(0,1)$, both $\Psc(k,n)$ and $\Psc(k-1,n-1)$ tend to one, whereas $\Sgc(k-1,n-1)$ tends to zero as $n\to\infty$. The stated limits follow.
\qed
\end{proof}

\begin{proof}[Theorem~\ref{thm:ac}]
By Lemma~\ref{lem:constraints}, an $\stratA\stratC$ profile with $|C|=\lambda$ is a Nash equilibrium exactly when \eqref{eq:I}, \eqref{eq:III}, \eqref{eq:V}, and \eqref{eq:VI} hold. Homogeneity makes these four conditions respectively the first inequality in \eqref{eq:ac-side}, the lower side of \eqref{eq:ac-window}, the upper side of \eqref{eq:ac-window}, and the second inequality in
\eqref{eq:ac-side}. This proves the fixed-cost characterization.

The window is nonempty exactly when $q[\Lam(\lambda) - \Lam(\lambda-1)] + r_1[\Psc(k,\lambda+1) - \Psc(k,\lambda)] \leq 0$. Using \eqref{eq:rec} and its exact-count analogue, $\Lam(\lambda) - \Lam(\lambda-1) = q\big[(r_2+v)\Sgc(k-2,\lambda-1) - v\Sgc(k-1,\lambda-1)\big]$ and $\Psc(k,\lambda+1) - \Psc(k,\lambda) = q\Sgc(k-1,\lambda)$. Substituting $\Sgc(k-1,\lambda) = q\Sgc(k-2,\lambda-1) + (1-q)\Sgc(k-1,\lambda-1)$ (the homogeneous exact-count recursion stated after \eqref{eq:rec}) and collecting terms gives exactly \eqref{eq:wac}. For $k = 1$ the left side of \eqref{eq:wac} vanishes, so the condition reduces to $qv \geq (1-q)r_1$. For $k \geq 2$, both binomial masses in \eqref{eq:wac} are positive. If $r_1+r_2+v=0$, \eqref{eq:wac} holds at equality but $\Lam(\lambda-1)=0$, so positive costs cannot be supported. Otherwise, its left side is positive, and \eqref{eq:wac} forces $qv > (1-q)r_1$. In the latter case, the ratio $\Sgc(k-1,\lambda-1)/\Sgc(k-2,\lambda-1) = \frac{\lambda-k+1}{k-1}\cdot\frac{q}{1-q}$ from \eqref{eq:binom-ratio} rearranges \eqref{eq:wac} to \eqref{eq:ac-lambda}.

It remains to characterize feasibility with positive costs. Necessity of \eqref{eq:wac} follows from window membership. Moreover, \eqref{eq:ac-side} and the upper side of the window give
\[
0<\beta\leq \alpha+\beta-r_1\Psc(k,\lambda)
\leq q\Lam(\lambda-1),
\]
so $\Lam(\lambda-1)>0$. Conversely, suppose \eqref{eq:wac} holds and $\Lam(\lambda-1)>0$. Set
\[
\varepsilon:=\tfrac12q\Lam(\lambda-1),\qquad
\beta:=\varepsilon,\qquad
\alpha:=r_1\Psc(k,\lambda)+q\Lam(\lambda-1)-\varepsilon.
\]
Then $\alpha,\beta>0$, $\alpha\geq r_1\Psc(k,\lambda)$, and $\beta\leq q\Lam(\lambda-1)\leq q\Pmf(\lambda-1)$. Also $\alpha+\beta=q\Lam(\lambda-1)+r_1\Psc(k,\lambda)$ is the upper endpoint of the nonempty window. Hence all four exact conditions hold. The case distinctions in (a)--(c) now follow from \eqref{eq:wac} and the algebra above.
\qed
\end{proof}

\begin{proof}[Theorem~\ref{thm:bc-hom}]
Because $A=\emptyset$, Lemma~\ref{lem:constraints} leaves exactly
\eqref{eq:II}, \eqref{eq:IV}, \eqref{eq:V}, and \eqref{eq:VI}. Homogeneity turns
them into the second, lower-window, third, and upper-window conditions of
\eqref{eq:bc-exact}, respectively. This proves necessity and sufficiency.

The window is consistent exactly when $\Pmf(\lambda)\leq\Pmf(\lambda-1)$. Expanding both sides and using the tail and exact-count recursions reduces this to
\[
(r_1+r_2+v)\Sgc(k-2,\lambda-1)
\leq (r_1+v)\Sgc(k-1,\lambda-1).
\]
For $k=1$, its left side is zero and its right side is nonnegative. For $k\geq2$, both masses in \eqref{eq:binom-ratio} are positive. If $r_1+v>0$, division and rearrangement give \eqref{eq:bc-lambda}; if $r_1+v=0$, the inequality reduces to $r_2\Sgc(k-2,\lambda-1)\leq0$, hence to $r_2=0$. Finally, substitute \eqref{eq:lam-phi} into the three $r_2$-dependent inequalities of \eqref{eq:bc-exact}. The remaining condition, $\alpha\leq r_1\Psc(k,\lambda)$, is independent of $r_2$; if it fails, no contribution reward supports the profile. If it holds, the other three inequalities are equivalent to
\begin{align*}
r_2 &\geq \frac{\beta-q(r_1+v)\Sgc(k-1,\lambda-1)}{q\Psc(k-1,\lambda-1)},
&
r_2 &\leq \frac{\beta-q(r_1+v)\Sgc(k-1,\lambda)}{q\Psc(k-1,\lambda)},\\
r_2 &\geq \frac{\alpha+\beta-r_1\Psc(k,\lambda)-qv\Sgc(k-1,\lambda-1)}
{q\Psc(k-1,\lambda-1)}.
\end{align*}
The lower bound on $\beta$ yields the upper reward bound; the upper bound on $\beta$ and the contributor's abstention constraint yield the two lower bounds. Their denominators are positive because $q\in(0,1)$ and $\lambda\geq k$. Reversing the algebra shows that these bounds, together with $\alpha\leq r_1\Psc(k,\lambda)$, are equivalent to \eqref{eq:bc-exact}.
\qed
\end{proof}

\begin{proof}[Theorem~\ref{thm:abc-hom}]
\emph{Necessity.} \eqref{eq:I} and \eqref{eq:II} give $\alpha \geq r_1\Psc(k,\lambda)$ and $\alpha \leq r_1\Psc(k,\lambda)$, hence $\alpha = r_1\Psc(k,\lambda)$. Under this knife-edge, $u_i(\stratA,s_{-i}) = v\Psc(k,\lambda) = (r_1+v)\Psc(k,\lambda) - \alpha = u_i(\stratB,s_{-i})$ for every $i\in A\cup B$, so every abstainer and bare registrant is indifferent between $\stratA$ and $\stratB$; in particular the constraints
\eqref{eq:III} and \eqref{eq:IV} coincide:
$\alpha + \beta \geq q\Lam(\lambda) + r_1\Psc(k,\lambda+1)$ becomes, after substituting $\alpha = r_1\Psc(k,\lambda)$ and $\Psc(k,\lambda+1) - \Psc(k,\lambda) = q\Sgc(k-1,\lambda)$, the inequality $\beta \geq q[\Lam(\lambda) + r_1\Sgc(k-1,\lambda)] = q\Pmf(\lambda)$. For contributors, \eqref{eq:V} with $\alpha = r_1\Psc(k,\lambda)$ simplifies to $\beta \leq q\Lam(\lambda-1)$, and \eqref{eq:VI} reads $\beta \leq q\Pmf(\lambda-1)$, which is implied since $\Lam \leq \Pmf$ pointwise. This yields
\eqref{eq:abc-cond}. Conversely, under \eqref{eq:abc-cond}, the same reductions verify all six equilibrium constraints. This proves the fixed-cost characterization.

\emph{Window consistency.} $q\Pmf(\lambda) \leq q\Lam(\lambda-1)$ expands to $r_2\Psc(k-1,\lambda) + (r_1+v)\Sgc(k-1,\lambda) \leq r_2\Psc(k-1,\lambda-1) + v\Sgc(k-1,\lambda-1)$; applying \eqref{eq:rec} and its exact-count analogue, $\Psc(k-1,\lambda) - \Psc(k-1,\lambda-1) = q\Sgc(k-2,\lambda-1)$ and $\Sgc(k-1,\lambda) = q\Sgc(k-2,\lambda-1) + (1-q)\Sgc(k-1,\lambda-1)$, gives $r_2 q \Sgc(k-2,\lambda-1) + (r_1+v)[q\Sgc(k-2,\lambda-1) + (1-q)\Sgc(k-1,\lambda-1)]
\leq v\Sgc(k-1,\lambda-1)$, i.e.\
$q(r_1+r_2+v)\Sgc(k-2,\lambda-1) \leq (qv - (1-q)r_1)\Sgc(k-1,\lambda-1)$, which is exactly \eqref{eq:wac}. The algebra of Theorem~\ref{thm:ac} gives the stated parameter restrictions for positive-cost coexistence, since $\Psc(k,\lambda)>0$ makes $\alpha>0$ equivalent to $r_1>0$. This also proves that the registration reward must be positive; indifference and non-strictness were shown above. For uniqueness of the supporting registration cost, fix the profile and all other parameters, including $\beta$. Then \eqref{eq:abc-cond} makes the set of positive supporting values of $\alpha$ empty unless $r_1>0$ and $q\Pmf(\lambda)\leq\beta\leq q\Lam(\lambda-1)$, and otherwise equal to the singleton $\{r_1\Psc(k,\lambda)\}$. Increasing $\alpha$ from this value violates \eqref{eq:II}, whereas decreasing it violates \eqref{eq:I}.

\emph{Positive-cost feasibility.} Suppose $r_1>0$ and \eqref{eq:wac} holds. Set $\alpha = r_1\Psc(k,\lambda)>0$. \eqref{eq:wac} makes the interval in
\eqref{eq:abc-cond} nonempty. Its upper endpoint is positive: for $k=1$, \eqref{eq:wac} and
$r_1>0$ imply $v>0$, while for $k\geq2$ the left-hand side of \eqref{eq:wac} is positive and therefore $qv>(1-q)r_1$, again implying $v>0$; in either case $\Lam(\lambda-1)>0$. Choose any positive $\beta$ in the interval. The verification above then shows all six constraints of Lemma~\ref{lem:constraints} hold for any partition with $|C| = \lambda$ and $A,B \neq \emptyset$.
\qed
\end{proof}

\appheading{Computation of Example~\ref{ex:abc}}
With $k = 2$, $q = 1/2$, $r_1 = r_2 = 1$, $v = 3$, $\lambda = 4$: $\Psc(2,4) = 11/16$; $q\Pmf(4) = \tfrac12[\Psc(1,4) + 4\Sgc(1,4)] = \tfrac12[\tfrac{15}{16} + 1] =
\tfrac{31}{32}$ and
$q\Lam(3) = \tfrac12[\Psc(1,3) + 3\Sgc(1,3)] = \tfrac12[\tfrac78 + \tfrac98] = 1$. Since $qv = 1.5 > 0.5 = (1-q)r_1$ and $\lambda_{\stratA\stratC} = 1 + 0.5\cdot 5/1 = 3.5 \leq 4$, Theorem~\ref{thm:abc-hom} applies.

\subsection{Heterogeneous selection}
\label{app:het}

\begin{proof}[Lemma~\ref{lem:equal}]
Suppose $e \in A$, $f \in C$, $q_e = q_f = \bar q$. Then \eqref{eq:III} for $e$ and
\eqref{eq:V} for $f$ imply
$\bar q \Lam(C) + r_1\Psc(k,C\cup\{e\}) \leq \alpha+\beta \leq \bar q\Lam(C_{-f}) + r_1\Psc(k,C)$. We show the leftmost quantity strictly exceeds the rightmost. Their difference $D$ expands via \eqref{eq:rec} and its exact-count analogue: since $\Lam(C) - \Lam(C_{-f}) = \bar q\big[(r_2+v)\Sgc(k-2,C_{-f}) - v\Sgc(k-1,C_{-f})\big]$ and $\Psc(k,C\cup\{e\}) - \Psc(k,C) = \bar q\,\Sgc(k-1,C) = \bar q\big[\bar q\,\Sgc(k-2,C_{-f}) + (1-\bar q)\Sgc(k-1,C_{-f})\big]$,
\[
D \;=\; \bar q^{2}(r_1+r_2+v)\,\Sgc(k-2,C_{-f})
\;+\; \bar q\big[(1-\bar q)r_1 - \bar q v\big]\,\Sgc(k-1,C_{-f}) \;>\; 0,
\]
because the first term is positive for $k \geq 2$: its probability factor is positive, and $r_1+r_2+v>0$ in any equilibrium with $C\neq\emptyset$ (otherwise every contributor strictly prefers $\stratA$). The second term is nonnegative by \hyperref[def:rd]{RD}; for $k = 1$ the first term vanishes and the second is positive by strict \hyperref[def:rd]{RD}. Hence $\alpha + \beta \geq \bar q\Lam(C) + r_1\Psc(k,C\cup\{e\}) > \bar q\Lam(C_{-f}) + r_1\Psc(k,C) \geq \alpha+\beta$, a contradiction.
\qed
\end{proof}

\begin{proof}[Lemma~\ref{lem:dom}]
Lemma~\ref{lem:equal} rules out $q_e=q_f$. Suppose for contradiction that $e\in A$, $f\in C$, and $q_e>q_f$. For $t \in [0,1]$ let $g(t) := t\,\Lam(C) + r_1\big[t\Psc(k-1,C) + (1-t)\Psc(k,C)\big]$, the right-hand side of \eqref{eq:III} for a hypothetical abstainer with selection probability $t$. Then $g'(t) = \Lam(C) + r_1[\Psc(k-1,C) - \Psc(k,C)] = \Lam(C) + r_1\Sgc(k-1,C) = \Pmf(C) > 0$ because $|C|\geq k$ and $r_1+r_2+v>0$ as shown in Lemma~\ref{lem:equal}; hence $g$ is strictly increasing. Moreover $g(q_f)$ minus the right-hand side of \eqref{eq:V} for $f$ equals the quantity $D$ of Lemma~\ref{lem:equal} with $\bar q = q_f$, which is positive. Hence
\[
\alpha+\beta \;\geq\; g(q_e) \;>\; g(q_f) \;>\;
q_f\Lam(C_{-f}) + r_1\Psc(k,C) \;\geq\; \alpha+\beta,
\]
a contradiction.
\qed
\end{proof}

\begin{proof}[Lemma~\ref{lem:monoC}]
For the first claim, let $C_{-ef}:=C\setminus\{e,f\}$ and expand both pools around this common set. For $w\geq0$, define $\Xi_w(x,S):=r_2\Psc(x,S)+w\Sgc(x,S)$; thus $\Xi_v(k-1,S)=\Lam(S)$ and $\Xi_{r_1+v}(k-1,S)=\Pmf(S)$. Applying both the tail and exact-count recursions after adding $e$ or $f$ to $C_{-ef}$ and cancelling the terms with coefficient $q_eq_f$ gives
\[
q_e\,\Xi_w(k-1,C_{-e}) - q_f\,\Xi_w(k-1,C_{-f})
\;=\; (q_e - q_f)\,\Xi_w(k-1,C_{-ef}) \;\leq\; 0,
\]
because $q_e<q_f$ and $\Xi_w\geq0$. Since $r_1\Psc(k,C)$ is common to both \eqref{eq:V} constraints, $f$ inherits
\eqref{eq:V} from $e$; and $\beta \leq q_e\Pmf(C_{-e}) \leq q_f\Pmf(C_{-f})$ gives
\eqref{eq:VI} for $f$. For the second claim, define
$h(t):=t\Lam(C)+r_1[t\Psc(k-1,C)+(1-t)\Psc(k,C)]$, the right-hand side of
\eqref{eq:III} apart from the common costs. Since $h'(t)=\Pmf(C)\geq0$ by
nonnegativity of rewards and probabilities, $h(q_e)\leq h(q_f)$, so the less-selectable abstainer inherits
\eqref{eq:III} from the more-selectable one.
\qed
\end{proof}

\begin{proof}[Theorem~\ref{thm:threshold}]
Lemmas~\ref{lem:equal} and \ref{lem:dom} show that no contributor can have a lower selection probability than an abstainer and that equal-probability users cannot split between $A$ and $C$. Hence every $\stratA\stratC$ equilibrium has the stated cutoff and, by Lemma~\ref{lem:constraints}, at least $k$ contributors. Lemma~\ref{lem:monoC} shows that
\eqref{eq:V}--\eqref{eq:VI} for the least-selectable contributor imply those constraints
for every contributor, while \eqref{eq:III} for the most-selectable abstainer implies it for every abstainer. Condition \eqref{eq:I} is independent of $q_i$. The boundary conditions are therefore necessary and sufficient for the candidate cutoff profile.
\qed
\end{proof}

\appheading{A nonvacuous instance for Theorem~\ref{thm:threshold}}
Let $n=2$, $k=1$, $v=0$, $r_1=r_2=1$, $(q_1,q_2)=(0.8,0.1)$, $\alpha=0.8$, and $\beta=0.5$. The profile $C=\{1\}$, $A=\{2\}$ satisfies strict \hyperref[def:rd]{RD}. Its abstainer inequalities are $0.8\geq0.8$ and $1.3\geq0.92$, while its contributor inequalities are $1.3\leq1.6$ and $0.5\leq1.6$.

\appheading{Failure under weak \hyperref[def:rd]{RD} when $k=1$}
Strictness at $k=1$ cannot be replaced by weak \hyperref[def:rd]{RD}. Let $n=2$, $k=1$, $q_1=q_2=1/2$, $r_1=v=1$, $r_2=0$, and $\alpha=\beta=1/2$. Weak \hyperref[def:rd]{RD} holds at equality, while $C=\{1\}$ and $A=\{2\}$ is an equilibrium: the two abstainer inequalities hold at equality, and the contributor inequalities have right-hand sides $1$ and $1$, respectively. Equal-probability users can therefore split between abstention and contribution commitment.

\begin{proof}[Theorem~\ref{thm:bc-het}]
The constraints \eqref{eq:II}, \eqref{eq:IV}, \eqref{eq:V}, \eqref{eq:VI} of Lemma~\ref{lem:constraints} must hold for the respective users. \eqref{eq:II} is type-independent, giving \eqref{eq:W3}. The right-hand side of \eqref{eq:IV} is $q_j\Pmf(C)$, increasing in $q_j$, so it suffices to check $j = \bar b$: \eqref{eq:W1}. For \eqref{eq:V} and \eqref{eq:VI}, the proof of Lemma~\ref{lem:monoC} shows that for $e, f \in C$ with $q_e < q_f$ one has $q_e\Lam(C_{-e}) \leq q_f\Lam(C_{-f})$ and $q_e\Pmf(C_{-e}) \leq q_f\Pmf(C_{-f})$; hence the minima are attained at $\underline c$, giving \eqref{eq:W2}, \eqref{eq:W4}. Define
\begin{align*}
L_1&:=\frac{\beta-q_{\underline c}(r_1+v)\Sgc(k-1,C_{-\underline c})}
{q_{\underline c}\Psc(k-1,C_{-\underline c})},\\
L_2&:=\frac{\alpha+\beta-r_1\Psc(k,C)}
{q_{\underline c}\Psc(k-1,C_{-\underline c})}
-\frac{v\Sgc(k-1,C_{-\underline c})}{\Psc(k-1,C_{-\underline c})},\\
U&:=\frac{\beta-q_{\bar b}(r_1+v)\Sgc(k-1,C)}{q_{\bar b}\Psc(k-1,C)}.
\end{align*}
The tails in these denominators are positive. Solving the four boundary conditions gives
\[
r_2\geq\max\{L_1,L_2\},\qquad r_2\leq U.
\]
Together with \eqref{eq:W3}, these bounds recover \eqref{eq:W1}--\eqref{eq:W4}, which imply every applicable constraint of Lemma~\ref{lem:constraints}. This proves sufficiency.
\qed
\end{proof}

\appheading{Verification of Example~\ref{ex:wrong-bc}}
Here $\Pmf(\{1\})=0.7$ and $\Pmf(\emptyset)=1.1$. The four conditions of Theorem~\ref{thm:bc-het} are $q_2\Pmf(\{1\})=0.35\leq0.4=\beta$, $\beta=0.4\leq0.44=q_1\Pmf(\emptyset)$, $\alpha=0.03\leq0.4=r_1\Psc(1,\{1\})$, and $\alpha+\beta=0.43\leq0.44=q_1\Lam(\emptyset)+r_1\Psc(1,\{1\})$.

\begin{proof}[Proposition~\ref{prop:restore}]
Suppose $j \in B$, $i \in C$, and $q_j \geq q_i$. Then $\beta \geq q_j\Pmf(C) \geq q_i\Pmf(C) \geq q_i\Pmf(C_{-i}) \geq \beta$ by \eqref{eq:IV}, pivotal non-dilution, and \eqref{eq:VI}. Moreover $\Pmf(C)>0$: condition \eqref{eq:II} and $\alpha>0$ force $r_1>0$, while $\Sgc(k-1,C)>0$. Hence $q_j>q_i$ makes the second comparison strict, and $\Pmf(C)>\Pmf(C_{-i})$ makes the third strict. Either case contradicts a chain that begins and ends at $\beta$. Applying this to a least-selectable contributor and a most-selectable bare registrant proves the weak order and both strictness clauses.
\qed
\end{proof}

\begin{proof}[Theorem~\ref{thm:abc-het}]
\eqref{eq:I} and \eqref{eq:II} force $\alpha = r_1\Psc(k,C)$, under which every
user in $A\cup B$ is indifferent between $\stratA$ and $\stratB$, and \eqref{eq:III} collapses to \eqref{eq:IV}: substituting $\alpha = r_1\Psc(k,C)$ into \eqref{eq:III}, $\beta \geq q_i\Lam(C) + r_1[\Psc(k,C\cup\{i\}) - \Psc(k,C)] = q_i[\Lam(C) + r_1\Sgc(k-1,C)] = q_i\Pmf(C)$, using \eqref{eq:rec}. The maximum over $A \cup B$ is at $\bar d$. For contributors,
\eqref{eq:V} becomes $\beta \leq q_i\Lam(C_{-i})$ (minimized at $\underline c$ by the
Lemma~\ref{lem:monoC} computation), which implies \eqref{eq:VI} since $\Lam \leq \Pmf$. This proves necessity.

Conversely, suppose \eqref{eq:abc-het} holds. Its equality satisfies both
\eqref{eq:I} and \eqref{eq:II}. The lower bound on $\beta$ implies
$\beta\geq q_i\Pmf(C)$ for every $i\in A\cup B$, so it satisfies \eqref{eq:IV} for $B$ and, by the displayed reduction, \eqref{eq:III} for $A$. The upper bound and the contributor monotonicity of Lemma~\ref{lem:monoC} give $\beta\leq q_i\Lam(C_{-i})$ for every $i\in C$, which is \eqref{eq:V} after the knife-edge equality. Since $\Lam(C_{-i})\leq\Pmf(C_{-i})$, \eqref{eq:VI} follows as well. Thus all six applicable deviation conditions hold, proving sufficiency. Non-strictness follows because each user in $A\cup B$ is indifferent between $\stratA$ and $\stratB$. For fixed values of the remaining parameters and a fixed profile, the required equality determines a single value of $\alpha$.

For the window comparison, choose $\bar b\in\arg\max_{j\in B}q_j$. Since $q_{\bar d}\geq q_{\bar b}$, $\Pmf(C)\geq0$, and $\Lam\leq\Pmf$, every value in the preparation-cost window of \eqref{eq:abc-het} also satisfies \eqref{eq:W1}--\eqref{eq:W2}. The converse can fail: take $n=3$, $k=1$, $q_i=1/2$ for every $i$, $r_1=r_2=1$, $v=0$, and $A,B,C$ each containing one user. Then $\Pmf(C)=3/2$, $\Pmf(\emptyset)=2$, and $\Lam(\emptyset)=1$. The window \eqref{eq:W1}--\eqref{eq:W2} is $[3/4,1]$, but the window in \eqref{eq:abc-het} requires $3/4\leq\beta\leq1/2$ and is empty, even at the required positive registration cost $\alpha=1/2$. Thus consistency of the former window does not imply three-action coexistence.
\qed
\end{proof}

\appheading{Verification of Example~\ref{ex:wrong-abc}}
Since $\alpha=r_1\Psc(1,\{3\})=0.5$, $\Pmf(\{3\})=1+11\cdot0.5=6.5$, and $\Lam(\emptyset)=1+10=11$, the boundary window in \eqref{eq:abc-het} is $q_2\Pmf(\{3\})=0.55\cdot6.5=3.575\leq4=\beta
\leq0.5\cdot11=5.5=q_3\Lam(\emptyset)$.

\subsection{Optimal reward design}
\label{app:design}

\begin{proof}[Theorem~\ref{thm:design}]
By Proposition~\ref{prop:single}, full contribution is a Nash equilibrium exactly when
\eqref{eq:allC} holds. Expanding $\Lam$ and $\Pmf$ gives two linear constraints on
$(r_1,r_2)$:
\begin{align*}
\Psc(k,n)\, r_1 + q\Psc(k-1,n-1)\, r_2 &\;\geq\; \Gamma_1 := \alpha+\beta - qv\,\Sgc(k-1,n-1),\\
q\,\Sgc(k-1,n-1)\, r_1 + q\Psc(k-1,n-1)\, r_2 &\;\geq\; \Gamma_2 := \beta - qv\,\Sgc(k-1,n-1).
\end{align*}
By \eqref{eq:outlay}, the outlay under all-$\stratC$ is $n\Psc(k,n)\,r_1 + nq\Psc(k-1,n-1)\,r_2$, exactly $n$ times the left-hand side of the first constraint. If $\Gamma_1\leq0$, then $\Gamma_2=\Gamma_1-\alpha<0$, so $(r_1,r_2)=(0,0)$ satisfies both constraints. Both outlay coefficients are strictly positive, so every nonzero nonnegative reward scheme has positive outlay. The zero-reward scheme is therefore the unique optimum, with outlay zero.

Now suppose $\Gamma_1>0$. Every feasible scheme costs at least $n\Gamma_1$, with equality if and only if the first constraint binds. On that binding line $r_2 = (\Gamma_1 - \Psc(k,n)r_1)/(q\Psc(k-1,n-1))$, the second constraint reads $r_1\big(\Psc(k,n) - q\Sgc(k-1,n-1)\big) \leq \Gamma_1 - \Gamma_2 = \alpha$, and $\Psc(k,n) - q\Sgc(k-1,n-1) = \Psc(k,n-1)$ by \eqref{eq:rec}; together with $r_1\geq0$ and $r_2 \geq 0$ this yields exactly the segment \eqref{eq:segment}. Its point with $r_1=0$ has $r_2=\Gamma_1/[q\Psc(k-1,n-1)]$. Combining the two cases gives \eqref{eq:opt-endpoint} with $\Gamma_1^+=\max\{\Gamma_1,0\}$.
\qed
\end{proof}

\begin{proof}[Corollary~\ref{cor:nofree}]
With $r_1=0$, for every $i$ and every $s_{-i}$,
\[
u_i(\stratB,s_{-i})=v\Psc(k,C_{-i})-\alpha
<v\Psc(k,C_{-i})=u_i(\stratA,s_{-i}),
\]
so $\stratB$ is strictly dominated and $B=\emptyset$ in equilibrium. If $v=0$ and $k\geq2$, \eqref{eq:wac} for a putative $\stratA\stratC$ equilibrium reads $qr_2\Sgc(k-2,\lambda-1)\leq0$. Its probability factor is positive for every $\lambda\geq k$; positive costs also force $r_2>0$ in any such equilibrium, because $r_1=v=0$ would otherwise give contributors negative payoff. Hence \eqref{eq:wac} fails. Theorem~\ref{thm:ac} excludes all partial contributor sets, leaving all-$\stratA$ and all-$\stratC$; only the latter has positive success probability.

For $k=1$, take $r_2=r_2^*+\varepsilon$ with $\varepsilon>0$. Since $\Psc(0,\cdot)=1$, $\Sgc(0,\lambda)=(1-q)^\lambda$, and $qr_2^*=\Gamma_1^+$, for every $1\leq\lambda\leq n-1$,
\[
q\Lam(\lambda)-(\alpha+\beta)
=q\varepsilon+(\Gamma_1^+-\Gamma_1)
+qv\big[(1-q)^\lambda-(1-q)^{n-1}\big]>0.
\]
Thus the lower endpoint of every $\stratA\stratC$ window exceeds $\alpha+\beta$. Replacing $(1-q)^\lambda$ by one in the same calculation gives $q(r_2+v)>\alpha+\beta$, violating the all-abstention condition in Proposition~\ref{prop:single}. Because $\stratB$ is dominated, every remaining profile is all-$\stratC$ or an $\stratA\stratC$ profile; the preceding arguments exclude the latter. Finally, increasing $r_2$ weakly relaxes both conditions in \eqref{eq:allC}, so the all-$\stratC$ equilibrium persists. It is therefore unique.
\qed
\end{proof}

\begin{proof}[Remark~\ref{rem:hetero-design}]
Abbreviate $P_N:=\Psc(k,N)$, $p_i:=\Psc(k-1,N_{-i})$, and $\sigma_i:=\Sgc(k-1,N_{-i})$. At all-$\stratC$, conditions
\eqref{eq:V}--\eqref{eq:VI} are, for every $i$,
\begin{align*}
P_Nr_1+q_ip_ir_2 &\geq \alpha+\beta-q_iv\sigma_i,\\
q_i\sigma_ir_1+q_ip_ir_2 &\geq \beta-q_iv\sigma_i.
\end{align*}
These inequalities are necessary and sufficient by Lemma~\ref{lem:constraints}, so they define the stated two-variable linear program. Since $q_ip_i>0$ for every $i$, sufficiently large $r_2$ is feasible; the positive outlay coefficients ensure that a minimum is attained. Assume balanced exposure, $nq_ip_i=\tilde Q$ for every $i$. Starting from any feasible scheme, reduce $r_1$ by $0\leq\delta\leq r_1$ and increase $r_2$ by $nP_N\delta/\tilde Q$; this leaves expected outlay unchanged. For each user $i$, the changes in the left sides of the two constraints are, respectively,
\[
-P_N\delta+q_ip_i\frac{nP_N\delta}{\tilde Q}=0,
\qquad
-q_i\sigma_i\delta+q_ip_i\frac{nP_N\delta}{\tilde Q}
=\delta\Psc(k,N_{-i})\geq0.
\]
The second equality uses $P_N-q_i\sigma_i=\Psc(k,N_{-i})$, which follows from \eqref{eq:rec}. Thus taking $\delta=r_1$ preserves feasibility and outlay and reaches $r_1=0$; applying this to an optimum proves the claim.

To identify when exposure is balanced, fix distinct $i,j$ and let $S:=N\setminus\{i,j\}$. Conditioning on the other user gives
\[
a_i=q_i\big[(1-q_j)\Psc(k-1,S)+q_j\Psc(k-2,S)\big],
\]
and the analogous expression for $a_j$ with $i,j$ interchanged. Subtraction yields
\[
a_i-a_j=(q_i-q_j)\Psc(k-1,S).
\]
If $k<n$, then $k-1\leq |S|$ and $\Psc(k-1,S)>0$, so all $a_i$ are equal if and only if all $q_i$ are equal. If $k=n$, then $a_i=\prod_{j\in N}q_j$ for every $i$, so exposure is balanced for arbitrary selection probabilities, including the case $n=k=1$.

For a concrete counterexample, let $n=2$, $k=1$, $(q_1,q_2)=(1/3,2/3)$, $\alpha=2$, $\beta=3$, and $v=0$. Then $P_N=7/9$, $p_1=p_2=1$, $\sigma_1=1/3$, $\sigma_2=2/3$, and $\tilde Q=1$. The two constraints of user~$1$ are
\[
\frac{7}{9}r_1+\frac{1}{3}r_2\geq5,
\qquad
\frac{1}{9}r_1+\frac{1}{3}r_2\geq3;
\]
for nonnegative rewards they imply the corresponding constraints of user~$2$. The objective is $(14/9)r_1+r_2$. Its feasible vertices are $(0,15)$, $(3,8)$, and $(27,0)$, with respective costs $15$, $38/3$, and $42$. More explicitly, along the lower feasible boundary the objective equals $15-7r_1/9$ for $0\leq r_1\leq3$ and $9+11r_1/9$ for $3\leq r_1\leq27$. It is therefore strictly decreasing before $(3,8)$ and strictly increasing after it. Hence the scheme with both rewards positive is the unique optimum, and every zero-registration-reward scheme supporting full contribution is strictly more expensive.
\qed
\end{proof}

\subsection{Social efficiency}
\label{app:eff}

\begin{proof}[Proposition~\ref{prop:opt}]
For part (i), moving every user in $B$ to $A$ leaves $C$ and hence $\Psc(k,C)$ unchanged while saving $\alpha|B|$ in \eqref{eq:welfare}; the welfare increase is strict because $\alpha>0$ and $B\neq\emptyset$.

For part (ii), \eqref{eq:rec} gives $\Psc(k,\lambda)-\Psc(k,\lambda-1)=q\Sgc(k-1,\lambda-1)$, and substitution yields the displayed increment. If every increment from $k$ through $n$ is nonnegative, then $W(n)\geq W(\lambda)$ for every $\lambda\in\{k-1,\ldots,n\}$. For $1\leq\lambda<k$, success is impossible and $W(\lambda)=-\lambda(\alpha+\beta)<0$, whereas $W(0)=0$. The final condition in the proposition gives $W(n)\geq0$. Consequently $W(n)$ is at least the welfare of every possible contributor count, so full contribution is socially optimal.
\qed
\end{proof}

\begin{proof}[Theorem~\ref{thm:pos}]
Theorem~\ref{thm:design} proves that \eqref{eq:opt-endpoint} is feasible, so all-$\stratC$ belongs to $\mathcal E$. If it is also a positive-welfare social optimum, then $W(\text{all-}\stratC)=\mathrm{OPT}>0$. No equilibrium can have welfare above $\mathrm{OPT}$, while this equilibrium attains it; hence $\max_{s\in\mathcal E}W(s)=\mathrm{OPT}$ and the definition gives $\mathrm{PoS}=1$.
\qed
\end{proof}

\begin{proof}[Theorem~\ref{thm:strong}]
Let $s$ be the all-$\stratC$ profile and let $s'$ result from a joint deviation by a nonempty coalition $T$. For user $i$ and a set $S\subseteq N\setminus\{i\}$ of other contributors, define
\begin{align*}
U^i_{\stratA}(S)&=v\Psc(k,S),\\
U^i_{\stratB}(S)&=(r_1+v)\Psc(k,S)-\alpha,\\
U^i_{\stratC}(S)&=-\alpha-\beta+q_i(r_1+r_2+v)\Psc(k-1,S)
 +(1-q_i)(r_1+v)\Psc(k,S).
\end{align*}
Adding a user to $S$ cannot decrease either Poisson-binomial tail probability. Because all coefficients are nonnegative, each $U^i_x(S)$ is therefore nondecreasing under set inclusion. Since $s$ is a Nash equilibrium, for every $i$ and $x\in\{\stratA,\stratB,\stratC\}$,
\[
U^i_x(N\setminus\{i\})\leq U^i_{\stratC}(N\setminus\{i\}).
\]
In the deviated profile let $S_i:=\{j\neq i:s'_j=\stratC\}$. Then $S_i\subseteq N\setminus\{i\}$, so for every $i\in T$,
\[
u_i(s')=U^i_{s'_i}(S_i)
\leq U^i_{s'_i}(N\setminus\{i\})
\leq U^i_{\stratC}(N\setminus\{i\})=u_i(s).
\]
Thus, in particular, no coalition can make all its members strictly better off, so $s$ is a strong equilibrium.
\qed
\end{proof}

\end{document}